\documentclass[10pt,twocolumn]{IEEEtran}
\usepackage{graphicx}
\usepackage{amsmath}
\usepackage{amssymb}
\usepackage[caption=false]{subfig}
\usepackage[noadjust]{cite}
\usepackage{float}
\usepackage{algorithm}
\usepackage{bibentry}
\usepackage{balance}
\usepackage{algorithm}
\usepackage{algorithmicx}
\usepackage{algpseudocode}
\usepackage{xcolor}

\graphicspath{{img/}}
\begin{document}

\title{Interference Avoidance in UAV-Assisted Networks: Joint 3D Trajectory Design and Power Allocation}

\author{Ali Rahmati$^*$, Seyyedali Hosseinalipour$^*$, Yavuz Yap{\i}c{\i}$^*$, Xiaofan He$^\dagger$, \.{I}smail G\"{u}ven\c{c}$^*$, \\ Huaiyu Dai$^*$, and Arupjyoti Bhuyan$^\ddagger$\\
$^*$Dept. of Electrical and Computer Engineering, North Carolina State University, Raleigh, NC\\$^\dagger$School of Electronic Information, Wuhan University, Wuhan, China\\
$^\ddagger$Idaho National Laboratory, Idaho Falls, ID\\
Email:{\tt \{arahmat, shossei3, yyapici, iguvenc, hdai\}@ncsu.edu,} \\{\tt xiaofanhe@whu.edu.cn, arupjyoti.bhuyan@inl.gov}\thanks{This work is supported in part through the INL Laboratory Directed Research \& Development (LDRD) Program under DOE Idaho Operations Office Contract DE-AC07-05ID14517.}
}

\maketitle

\begin{abstract}
The use of the unmanned aerial vehicle (UAV)  has been foreseen as a promising technology for the next generation communication networks. Since there are no regulations for UAVs deployment yet, most likely they form a network in coexistence with an already existed network. In this work, we consider a transmission mechanism that aims to improve the data rate between a terrestrial base station (BS) and user equipment (UE) through deploying multiple UAVs relaying the desired data flow. Considering the coexistence of this network with other established communication networks, we take into account the effect of interference, which is incurred by the existing nodes. Our primary goal is to optimize the three-dimensional (3D) trajectories and power allocation for the relaying UAVs to maximize the data flow while keeping the interference to existing nodes below a predefined threshold. An alternating-maximization strategy is proposed to solve the joint 3D trajectory design and power allocation  for the relaying UAVs. To this end, we handle the information exchange within the network  by resorting to spectral graph theory and subsequently address the power allocation through convex optimization techniques. Simulation results show that our approach can considerably improve the information flow while the interference threshold constraint is met.
\end{abstract}
\vspace{-1.5mm}

\vspace{-5mm}

\section{Introduction}
 The utilization of unmanned aerial vehicles (UAVs) has recently become a practical approach for a variety of mission-driven applications including border surveillance, natural disaster aftermath, monitoring, search and rescue, and purchase delivery \cite{hayat2016survey,rahmati2019energy}. Owing to the low acquisition cost of UAVs as well as their fast deployment and efficient coverage capabilities, UAV-assisted wireless communications has attracted lots of interest recently \cite{mag,7577063,mag2}. Specifically, the 3D mobility feature of UAVs and the coexistence of relaying UAVs with other existing communication networks (e.g., cellular networks) has led to new design challenges and opportunities in these networks~\cite{YOUNIS2008621}. 

The current literature on UAV-assisted wireless communications can be split into two categories according to the dynamics of UAVs during the data transmission. The \textit{static} frameworks assume that the geometric locations of all the network nodes (involving UAVs) remain unchanged during the data transmission \cite{zhan2006wireless,8424236}, while the \textit{dynamic} scenarios assume that UAVs can be mobile in any phase of the data transmission \cite{faqir2018energy, 8116613, zhang2018joint, zeng2016throughput}. In \cite{zhan2006wireless}, the optimal deployment of a static UAV is considered to improve the average data rate between two obstructed access points under a symbol error rate  constraint. Considering multiple static UAVs, optimal UAV locations are derived in \cite{8424236} through maximizing the data rate in single link multi-hop and multiple links dual-hop relaying schemes. As a follow up work for \cite{8424236},  we studied the optimal position planning of UAV relays considering both the existence of a single UAV and multiple UAVs between the transmitter and the receiver, which coexist with a major source of interference in the environment \cite{hosseinalipour2019interference}.

In the context of \textit{dynamic} UAV-assisted wireless communications,  \cite{faqir2018energy} considers the joint optimization of the propulsion and the transmission energies for relaying UAVs. An optimal control problem is accordingly formulated based on energy minimization considering dynamic models for both transmission and mobility. In \cite{8116613}, the optimum altitude of a relaying UAV is derived so as to maximize the reliability of the system, which is measured by the total power loss,
the overall outage, and the overall bit error rate. A UAV-assisted communication scheme is proposed in \cite{zhang2018joint}, where the UAV trajectory, and the transmit power of both the UAV and the mobile device are obtained via minimizing
the outage probability.

In this work, we consider a communication scenario where multiple dynamic UAVs are deployed to improve the data flow between a terrestrial base station (BS) and a desired user equipment (UE). As a major difference from the existing  literature, we take into account the interference due to the interaction between the newly deployed UAVs and the existing network nodes (e.g., neighboring BSs, small cells, jammers) in a dynamic setting.  We take  advantage of the inherent feature of the UAVs, i.e., their mobility, in order to evade from the interference of the co-existed network. That means the UAVs can reconfigure their locations to decrease the effect of mutual interference. Even though considering just 3D trajectory design can help the UAV network to avoid the unwanted interference from the co-existed network, it can be problematic for the co-existed network. In particular,  given the UAVs transmission powers, there is no guarantee that the interference constraint is met at the  co-existed primary network with the 3D trajectory design solely.  Thus, our goal is to jointly optimize the  trajectory and the power allocation of the relaying UAVs in 3D space  while keeping the interference generated by the UAVs to the existing nodes below a threshold. Since joint optimization of UAV 3D trajectories and powers is highly nontrivial, we propose a mechanism based on the alternating-maximization approach~\cite{bezdek2002some}. In our method, the  optimization problem is decomposed into two sub-problems where 3D trajectories are obtained using spectral graph theory tools while the power allocation problem is solved using convex optimization techniques. 
 

\vspace{-3mm}
\section{System Model}\label{sec:system}


\subsection{Communication Scenario}

\begin{figure}[!t]
	\includegraphics[width=0.44\textwidth]{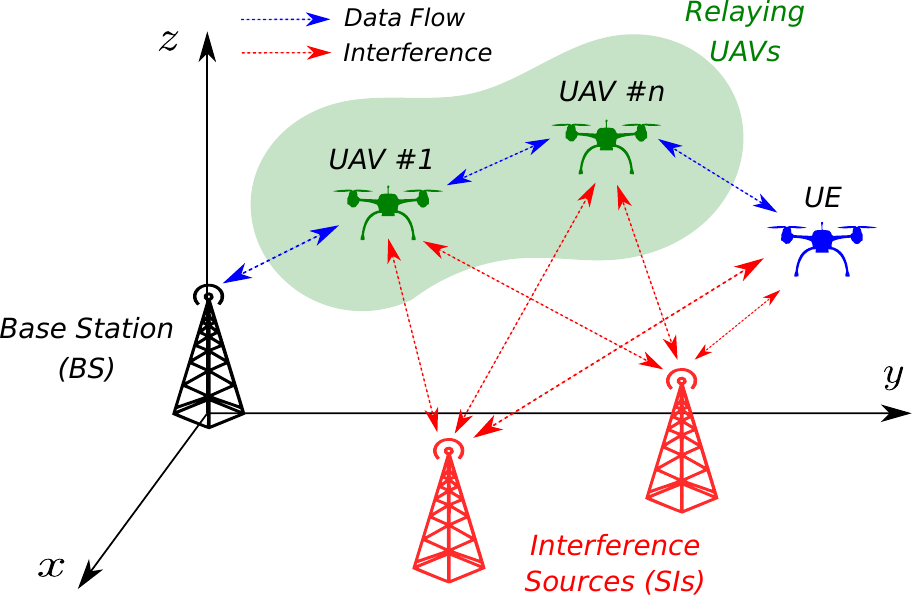}
	\centering
	\caption{System model for the communications scenario where the desired UE is also a flying UAV.}
	\label{fig:system}
\end{figure}

\noindent We consider a scenario where a terrestrial BS and a UE aim to engage in signal transmission. The UE is either  on the ground (e.g., a moving vehicle, pedestrian) or  in the air (e.g., a UAV), as shown in Fig.~\ref{fig:system}. To improve the data rate, we consider employing multiple UAVs relaying signal transmission between the BS and the UE. We also take into account any interference coming from the existing network (e.g., neighboring BSs, small cells, or malicious jammers), and describe the corresponding transmitters as \textit{sources of interference (SIs)}. We assume that the SIs can be detected together with their transmission parameters using existing sensing methods in the literature (e.g., \cite{tavana2017cooperative}). Moreover, we assume that the BS, the UE, the UAVs and the SIs are functioning as both transmitters and receivers, i.e.,  transceivers, and thus can involve in both uplink and downlink of their own respective networks. 

We adopt time-division multiple access (TDMA) to schedule the relaying UAVs so that their  transmissions do not collide with each other (i.e., the transmissions happen in the same frequency band but at different time slots). Our  goal is to 
\textit{obtain the 3D trajectories of the relaying UAVs along with the power allocation to maximize the data rate flow between the BS and the UE}. In addition to the communication-related applications, another use case for this scenario is in aerial wireless sensor networks involving several UAVs equipped with suitable sensors and radio devices, which fly over an area of interest to sense and collect data. 

In our setting, $\mathcal{N}$ describes the set of $N$ nodes in our network, which consists of the terrestrial BS (denoted by node $s$), the desired UE (denoted by node $d$), and the relaying UAVs. In addition, $\mathcal{M}$ stands for the set of $M$ separate SIs. The geometric location of any node in either of these sets is denoted by $\textbf{r}_n \,{=}\, (x_n,y_n,z_n) \,{\in}\, \mathbb{R}^3$ such that $n\,{\in}\, \mathcal{M}$ or $n\,{\in}\, \mathcal{N}$. In the subsequent sections, we distinguish between the aerial and terrestrial nodes, which are represented by $\mathcal{A} \,{=}\, \mathcal{N} \,{\backslash}\, \{s, d\}$ and $\mathcal{G} \,{=}\, \mathcal{M} \,{\cup}\, \{s, d\}$, respectively, and assume that the UE is placed on the ground. For the case of flying UEs, these sets are denoted by $\mathcal{A} \,{=}\, \mathcal{N} \,{\backslash}\, \{s\}$ and $\mathcal{G} \,{=}\, \mathcal{M} \,{\cup}\, \{s\}$. 
\vspace{-3mm}
\subsection{A2A and A2G Channel Models}
 In this section, we discuss the air to air (A2A) and the air to ground (A2G) channel models under consideration. 
In this work, we consider transmission through the line of sight (LoS) path only (e.g., \cite{8424236}), and adopt a widely-used path-loss model provided by the International Telecommunication
Union (ITU). The  path-loss for the A2A link is therefore given (in dB scale) as
\begin{equation}
    \textrm{PL}^{\textrm{A2A}}(d_{i,j})=\alpha_1 10 \log_{10} d_{i,j}+ \eta_1 ,
\end{equation}
where $\alpha_1$ is the path-loss exponent, $d_{i,j}$ is the distance between the UAVs $i$ and $j$, and $\eta_1$ is the path-loss associated with the reference LoS distance of $1\,\text{m}$. Similarly, the A2G channel between any terrestrial node and the aerial node can be described by the path-loss expression, given by
\begin{equation}
    \textrm{PL}^{\textrm{A2G}}(d_{i,j})=\alpha_2 10 \log_{10} d_{i,j}+ \eta_2 ,
\end{equation}
where $\alpha_2$ is the path-loss exponent, and $\eta_2$ is the path-loss at the reference distance. Note that in the free space, the path-loss exponents would be $\alpha_\ell\,{=}\,2$, and the path-loss at the reference distance would become $\eta_\ell \,{=}\, 10\log_{10} \,(\frac{4\pi f_{\rm c}}{c})^2$, where $f_{\rm c}$ is the carrier frequency, $c\,{=}\,3\times 10^8\,\text{m/s}$ is the speed of  light, and $\ell \,{\in}\, \{1,2\}$.
In the literature, the A2G channel is discussed to be close to a free-space
propagation scenario. Adopting \cite{ahmed2016importance}, we consider a larger path-loss exponent for the A2G channels  compared to the A2A channels, where these parameters are numerically presented in Section~\ref{sec:results}. Considering the impact of small-scale fading, the overall complex channel gain for nodes $i$ and $j$ involved in either the A2A or A2G communications becomes
\begin{equation}
h_{i,j}=\begin{cases}
\displaystyle\frac{g_{i,j}}{\sqrt{\textrm{PL}^{\textrm{A2A}}(d_{i,j})}},& \textrm{if the link is A2A}, \\
\displaystyle\frac{g_{i,j}}{\sqrt{\textrm{PL}^{\textrm{A2G}}(d_{i,j})}}, & \textrm{if the link is A2G},
\end{cases}
\end{equation} 
where $g_{i,j}$ is the small-scale fading gain, which is zero-mean complex Gaussian with unit variance. As a final remark, we assume  channel reciprocity for all the links under consideration so that the link $(i,j)$ from the node $i$ to node $j$ is the same as the link $(j,i)$ in the opposite direction.

\section{Network Topology and Problem Formulation}

\subsection{Graph Representation of the Network} \label{sec:sir_laplacian}

We assume that   the interference coming from the SIs is much stronger than the noise. We therefore take into account the SIR as the performance metric, which is defined for the transmission link from the node $i$ to $j$ as follows:
\begin{align}\label{eq:sir}
\textrm{SIR}_{i,j} = \frac{P_i \,  |h_{i,j}|^2}{\sum\limits_{m\in\mathcal{M}} {P}_m^\textrm{J} \,|h_{m,j}|^2+\chi\sum\limits_{k \in {\mathcal{Q}}_{i,j}}u(d_{j,k}/r_{\textrm{int}})},
\end{align}
where  $\mathcal{Q}_{i,j} \,{=}\, \mathcal{N} \,{\backslash}\, \{i,j\}$, $P_i$ is the transmit power of node $i$ in the  primary network (i.e., $i\,{\in}\,\mathcal{N}$), 
and ${P}_m^\textrm{J}$ is the transmit power of node $m$ in the SI network (i.e., $m\,{\in}\,\mathcal{M}$). 
The second term in the denominator of \eqref{eq:sir} is considered to guarantee a safety separation between any of the UAVs and other nodes in the primary network (i.e., the BS, the desired UE, or the other UAVs) so as to preserve a proper flight performance~\cite{weibel2004safety}. In this representation, $\chi$ stands for the importance of this safety precaution, and $u$ is the \textit{smoothed} step function given by \cite{6807812}:
\begin{align}\label{eq:smooth_step}
u(y) = \zeta \frac{\textrm{exp}(- \kappa y - \log y_0)}{1+\textrm{exp}(- \kappa y - \log y_0)},    
\end{align}
where $y_0$ is an arbitrarily  small positive number, and $\zeta$ and $\kappa$ are design parameters. 
We first define a directed flow graph $G \,{=}\,(\mathcal{N},\mathcal{E})$, in which each edge has the capacity $a_{i,j}$ and  $\mathcal{E}$ denotes the set of available edges in the network. We assume that the UAVs form a line graph which is a commonly used assumption in multi-hop relay networks~\cite{8424236}. We formulate the information exchange in this single-source and single-destination network using a \textit{single-commodity maximum flow problem}, for which the task is to determine the maximum amount of flow, i.e., the maximum average transmission rate, between the BS and  the desired UE.

We represent the information exchange among different nodes of the primary wireless network using a \textit{graph}. In this representation, existence of any edge between different nodes of the network is described by a nonzero entry in the \textit{generalized adjacency matrix} $\mathbf{A}$~\cite{chung1997spectral}.  More specifically, the weight of each edge in the line graph is determined according to $\textrm{SIR}_{i,j}$ defined in \eqref{eq:sir}. Note that we assume that the UAVs are operating as transceivers while relaying, and the BS and the desired UE are also capable of both transmitting and receiving (depending on uplink/downlink mode). We therefore can define the average transmission rate as the arithmetic mean of the data rates  in the forward and backward directions  for each pair of nodes. The generalized adjacency matrix is accordingly defined as $\mathbf{A} \,{=}\,[a_{i,j}]_{\{i,j\}{=}1}^{N}$, where $a_{i,j}$ is the average transmission rate between nodes $i$ and $j$, given by:
\begin{equation}\label{eq:edge}
a_{i,j}=\begin{cases}
\frac{1}{2}{B} \Big({\log_2(1{+}\textrm{SIR}_{i,j})}{+}{\log_2(1{+}\textrm{SIR}_{j,i})} \Big), & i {\ne} j,\\ 
0, & i{=}j,
\end{cases}
\end{equation}
where ${B}$ is the transmission bandwidth of the network. Note that $\textrm{SIR}_{i,j}$ is, in general, not equal to $\textrm{SIR}_{j,i}$, in part, due to the unbalanced deployment of SIs. 
 We further define 
the \textit{generalized degree matrix} of the network as $\mathbf{D}=\textrm{diag}\{\beta_1, ..., \beta_N\}$, where $\beta_i= \sum_{\{j|j \ne i\}}a_{i,j}$ denotes the number of edges (i.e., degree) attached to each node. Finally, the \textit{Laplacian matrix} of the network graph is then given by $\mathbf{L} \,{=}\, \mathbf{D} \,{-}\, \mathbf{A}$.

\subsection{Maximum Flow  Problem  \label{sec:optimization_problem}}

{\color{black} In this section, we formulate the optimization problem for the overall network, where the  goal is to maximize the data flow $\textrm{R}_{\textrm{s} \leftrightarrow \textrm{d}}$ between the terrestrial BS and the desired UE with the help of relaying UAVs in the presence of  SIs. } 
For each link $(i,j) \,{\in}\, \mathcal{E}$, let $f_{i,j}$ be the associated flow such that $0 \,{\le}\, f_{i,j} \,{\le}\, a_{i,j}$. The desired optimization problem is therefore given as follows
\begin{IEEEeqnarray}{rl}
\max_{\substack{P_i, \textbf{r}_i \\ \forall i \in \mathcal{Q}_{s,d}}}
&\qquad \textrm{R}_{\textrm{s} \leftrightarrow \textrm{d}} \;\,{=}\, \!\!\!\sum\limits_{j:(s,j) \in \mathcal{E}} f_{s,j}   \label{eqn:joint_opt_1}\\
\text{s.t.}
&\qquad \!\!\!\sum\limits_{i:(i,j) \in \mathcal{E}} f_{i,j} - \!\!\!\sum\limits_{l:(j,l) \in \mathcal{E}} f_{j,l} \,{=}\, 0, \; \forall  j \in \mathcal{Q}_{s,d}, \IEEEyessubnumber\label{eqn:joint_opt_2}\\
&\qquad 0 \le f_{i,j} \le a_{i,j}, ~ \forall (i,j) \in \mathcal{E},  \IEEEyessubnumber\label{eqn:joint_opt_3}\\
&\qquad   P_i |h_{i,j}|^2 \le \textrm{I}^{\textrm{max}}_{j}, \; \forall i \in \mathcal{N}, j \in \mathcal{M}, \IEEEyessubnumber\label{eqn:joint_opt_4}\\
&\qquad P_i \le \textrm{P}_{\textrm{max}}, \; \forall i \in \mathcal{N}, \IEEEyessubnumber \label{eqn:joint_opt_5}
\end{IEEEeqnarray}
where $P_i$ and $\textbf{r}_i$ stand for the power and geometrical location of the $i$th UAV, respectively, $\textrm{P}_{\textrm{max}}$ is the maximum transmit power of each UAV, and $\textrm{I}^{\textrm{max}}_{j}$ represents the predefined interference threshold for the $j$th SI. Note that the location vector $\textbf{r}_i$ involves in $a_{i,j}$ through the complex channel gain $h_{i,j}$ by \eqref{eq:sir} and \eqref{eq:edge}. In addition, \eqref{eqn:joint_opt_2} is due to the assumption of balanced flows for all the nodes except the  source and the destination. Moreover, \eqref{eqn:joint_opt_4} satisfies the condition that the interference produced by the each UAV at any SI is always less than a predefined threshold $\textrm{I}^{\textrm{max}}_{j}$ for the $j$th SI. 

It is worth mentioning that  given the UAVs transmission powers, there is no guarantee that the interference constraint is met at the  co-existed primary network with the 3D trajectory design solely~\cite{rahmati2019dynamic}. On the other hand, assuming fixed locations for UAVs, solely optimizing the UAV transmission  powers leads to a poor performance at the UAV relay network. Thus, joint power allocation and 3D trajectory design is necessary to obtain the satisfactory performance for both networks, which is therefore very complicated to solve. In the following, we propose to decompose the overall optimization of~\eqref{eqn:joint_opt_1} into two sub-problems using the alternating-optimization approach~\cite{bezdek2002some}. In the proposed strategy, we first solve the problem of 3D trajectory optimization for a given set of transmit powers (i.e., $P_i, \forall i \in \mathcal{N}$), and then the power allocation problem is solved for the given set of UAV locations computed beforehand. {\color{black}These recursions continue till a satisfactory level of performance is obtained.} 

\section{Joint 3D Trajectory Design and Power Allocation}\label{sec:interference_type}


\noindent In this section, we consider the solution of the 3D trajectory and power allocation problem for the UAVs using the alternating-projection strategy introduced in Section~\ref{sec:optimization_problem}.

\subsection{{\color{black}  3D Trajectory Design}}
Given the transmit power values of the UAVs, we first attempt to solve the optimization problem in \eqref{eqn:joint_opt_1} for the 3D trajectories. In this case, the problem reduces to a maximum flow problem with respect to the locations, which is given by:
\begin{IEEEeqnarray}{rl}
\max_{\substack{\textbf{r}_i \\ \forall  i \in \mathcal{Q}_{s,d}}}
&\qquad \textrm{R}_{\textrm{s} \leftrightarrow \textrm{d}} \;\,{=}\, \!\!\!\sum\limits_{j:(s,j) \in \mathcal{E}} f_{s,j}   \label{eqn:unin_loc_opt_1}\\
\text{s.t.}
&\qquad \eqref{eqn:joint_opt_2},  \eqref{eqn:joint_opt_3}, \nonumber 
\end{IEEEeqnarray}
Note that the maximum flow problem in \eqref{eqn:unin_loc_opt_1} can be solved for a given UAV location using the well-known max-flow- min-cut theorem~\cite{ford2015flows}. The achievable maximum flow of the network is equal to single flow min-cut of the underlying network given by:
\begin{equation}\label{fflow}
     \textrm{R}^{\textrm{max}}_{\textrm{s} \leftrightarrow \textrm{d}}= \min_{\substack{\{S:v_s\in S, v_d \in \bar S\}}} \sum\limits_{i\in S, j \in \bar S} a_{i,j}
\end{equation}
The maximum flow in \eqref{fflow} can be obtained by the Ford-Fulkerson algorithm \cite{ford2015flows}.
The more challenging task, however, is to design the trajectories (i.e., moving directions) of  each UAV in the 3D space so as to maximize the information flow between the BS and the desired UE. 


In order to move towards the maximum flow trajectory, we use \textit{Cheeger constant} or\textit{ isoperimetric number} of the graph which provides numerical measure on how well-connected our multi-node primary wireless network is \cite{chung1997spectral}. Assuming that $\mathcal{L} \,{=}\, \mathbf{D}^{-1/2}\mathbf{L} \mathbf{D}^{-1/2}$ is the normalized Laplacian matrix, the Cheeger constant is given by \cite{chung1997spectral}:
\begin{equation}\label{eq:cheeger}
\displaystyle h(\mathcal{L}) = \underset{S}{\text{min}} \frac {\sum_{i \in S, j \in \bar S}a_{i,j}}{\text{min}\{ |S|, |\bar S|\}},
\end{equation}
where $S \,{\subset}\, \mathcal{N}$ is a subset of the nodes, {\color{black} $\bar S \,{=}\, \mathcal{N} \,{-}\, S$}, and $|S|$ is the cardinality of set S. 
Note that the original definition of the Cheeger constant $h(\mathcal{L})$ considers all the nodes in the network with equal importance. Since the maximum flow of the network for a given source-destination pair depends on the individual link capacities, the weighted version of the Cheeger constant appears as a promising solution to overcome this drawback. In particular, the original Cheeger constant blindly aims at improving the weakest link in the network and may fail to emphasize the desired flow associated with a particular source-destination pair. 
We therefore need to distinguish between the BS and the desired UE from the UAV nodes, for which the weighted Cheeger constant comes as a remedy, and is given as \cite{6807812}:
\begin{equation}\label{eq:weighted_cheeger}
h_{\mathbf{W}}(\mathcal{L})=\underset{S}{\text{min}} \frac {\sum_{i \in S, j \in \bar S}a_{i,j}}{\text{min}\{ |S|_{\mathbf{W}}),(|\bar S|_{\mathbf{W}})\}},
\end{equation}
where $|S|_{\mathbf{W}} \,{=}\, \sum_{i \in S}w_i$ is the weighted cardinality, and $w_i \,{\ge}\, 0$ is the weight of the node $i$ which is adopted to emphasize   the flow between the BS and the desired UE. 
The weighted Laplacian matrix is accordingly given by:
\begin{equation}\label{eq:weighted_laplacian}
\mathcal{L}_\mathbf{W}=\mathbf{W}^{-1/2}\mathcal{L} \mathbf{W}^{-1/2},
\end{equation}
where $\mathbf{W} \,{=}\,{\rm diag} \{w_1, ...,w_n\}$. To achieve our goal, the ideal choice is to consider the weights
for the source and destination as 1, and those for the UAVs
as 0. However, in practice, we consider small non-zero weights for the UAVs as it may cause numerical issues. The weighted second smallest eigenvalue $\lambda_2({\mathcal{L}_\mathbf{W}})$ can be defined as
\begin{equation}
\lambda_2(\mathcal{L}_\mathbf{W})= \underset{\textbf{v} \ne \textbf{0}, \textbf{v} \perp \mathbf{W}^{1/2}\textbf{1}}{\text{min}} \frac{ \langle \mathcal{L}_\mathbf{W} \textbf{v}, \textbf{v} \rangle}{\langle \textbf{v}, \textbf{v}\rangle}. 
\end{equation}
It is shown in \cite{6807812}, the following weighted Cheeger's inequalities hold
\begin{equation}
\lambda_2(\mathcal{L}_\mathbf{W})/2 \le h_\mathbf{W}(\mathcal{L}_\mathbf{W}) \le \sqrt{2 \delta_{\textrm{max}} \lambda_2(\mathcal{L}_\mathbf{W})/w_{\textrm{min}}} \,,
\end{equation}
where $\delta_{\textrm{max}}$ is the maximum node degree {\color{black}(i.e., maximum number of edges attached to any node)}, 
and $w_{\textrm{min}}= \min_i w_i$. In practice, since computing the Cheeger constant is difficult, we try to maximize  $\lambda_2({\mathcal{L}_\mathbf{W}})$ as an alternative. When $\lambda_2(\mathcal{L}_\mathbf{W})$ gets larger values, the lower bound of the weighted Cheeger constant inequality increases, which improves the connectivity of the overall network.  The UAVs should therefore adjust their geometrical locations in order to maximize $\lambda_2({\mathcal{L}_\mathbf{W}})$, and hence $h_{\mathbf{W}}(\mathcal{L})$.

As a result, each UAV should move along the \textit{spatial} gradient of the weighted algebraic connectivity $\lambda_2({\mathcal{L}_\mathbf{W}})$ to maximize it. Given the instantaneous location of the $i$th UAV, its spatial gradient along $x$-axis is given as follows:
\begin{align}
\frac{\partial \lambda_2({\mathcal{L}_\mathbf{W}}) }{\partial x_i} & =    {\mathbf{x}^f}^T \frac{\partial ({{\mathcal{L}_\mathbf{W}}}) }{\partial x_i} {\mathbf{x}^f} \label{eq:spatial_gradient_unintended_1}\\
&= \sum_{\{p,q:p \sim q \}}  \left[\frac{x_p^f}{\sqrt{w_p}}-\frac{x_q^f}{\sqrt{w_q}}\right]^2\frac{\partial a_{p,q}}{\partial x_i}, & 
\label{eq:spatial_gradient_unintended_2}
\end{align}
where $\mathbf{x}^f$ is the Fiedler vector being the eigenvector corresponding to the second smallest eigenvalue {\color{black}$\lambda_2({\mathcal{L}_\mathbf{W}})$}, ${x}^f_k$ is the $k$th entry of $\mathbf{x}$ with $k\,{\in}\,{p,q}$, and $p\,{\sim}\,q$ means that
the nodes $p$ and $q$ are connected. In order to compute \eqref{eq:spatial_gradient_unintended_2}, we need to compute $\frac{\partial a_{p,q}}{\partial x_i}$, which is given by:
\begin{align}\label{eq:partial_derivative}
\frac{\partial a_{p,q}}{\partial x_i} &= \textrm{B}
\left[ \frac{1}{1{+}\textrm{SIR}_{p,q}}\frac{\partial \textrm{SIR}_{p,q}}{\partial x_i}  + \frac{1}{1{+}\textrm{SIR}_{q,p}}\frac{\partial \textrm{SIR}_{q,p}}{\partial x_i} \right],
\end{align}
which is $0$ for $p\,{=}\,q${\color{black}, or $i \,{\notin}\,\{p,q\}$}. The partial derivative of SIR with respect to $x_i$ in \eqref{eq:partial_derivative} can be computed using \eqref{eq:sir} together with the geometrical relations between $x_i$ and the complex channel gain $h_{i,j}$ presented in Section~\ref{sec:system}. The update in the location of the $i$th UAV along the $x-$axis is then given by:
\begin{align}\label{eq:loc_update_x}
    x_i(t+1)=x_i(t)+{\rm d}t \frac{\partial \lambda_2({\mathcal{L}_\mathbf{W}}) }{\partial x_i(t)},
\end{align}
where $t$ stands for the discrete time, or, equivalently, the iteration number. Note that a similar procedure can be pursued to find the spatial gradients of $\lambda_2({\mathcal{L}_\mathbf{W}})$ along $y-$ and $z-$axis, and update the respective coordinates. 
Here, we can consider both 2D and 3D trajectory designs for interference avoidance of the UAVs. Considering a 2D trajectory design in $xy$-plane, the update of the locations in $x$ and $y$ should be according \eqref{eq:loc_update_x}, while for $z$ axis, we do not consider any change in the location of UAVs in $z$ direction. For 3D trajectory design, we consider the location update in all three  directions.



\subsection{Power Allocation Optimization}

We now focus on the power allocation problem, and solve the optimization problem in \eqref{eqn:joint_opt_1} to find optimal power allocation for a given set of the UAV locations. In this case, the corresponding optimization problem is given by: 
\begin{IEEEeqnarray}{rl}
\max_{\substack{P_i \\ \forall i \in \mathcal{Q}_{s,d}}}
&\qquad \textrm{R}_{\textrm{s} \leftrightarrow \textrm{d}} \;\,{=}\, \!\!\!\sum\limits_{j:(s,j) \in \mathcal{E}} f_{s,j}   \label{eqn:unin_pow_opt_1}\\
\text{s.t.}
&\qquad \!\eqref{eqn:joint_opt_2},  \eqref{eqn:joint_opt_3}, \eqref{eqn:joint_opt_4}, \eqref{eqn:joint_opt_5}. \nonumber 
\end{IEEEeqnarray}
We assume that the UAVs transfer the data by forming a multi-hop single link network topology, i.e., a line graph, between the BS and the UE. 
This is a reasonable assumption as the major information exchange happens between neighboring UAVs, not between the UAVs away by more than a single hop.
In this case, the maximum  transmission rate of the network is determined by the hop with the minimum transmission rate.
More specifically, the power allocation problem can be equivalently given by  
\begin{IEEEeqnarray}{rl}
\max_{\substack{P_i , \,\forall i \in \mathcal{Q}_{s,d}}} \;  \min_{\substack{ \forall \, j \in \mathcal{Q}_{s,d}}}
&\qquad a_{i,j}  \label{eqn:unin_pow_opt_2}\\
\text{s.t.}
&\qquad \eqref{eqn:joint_opt_4}, \eqref{eqn:joint_opt_5}. \nonumber 
\end{IEEEeqnarray}
In order to have a more tractable problem, \eqref{eqn:unin_pow_opt_2} can be reformulated as follows
\begin{IEEEeqnarray}{rl}
\max_{\substack{P_i , \,\forall i \in \mathcal{Q}_{s,d}}} & \qquad \eta \label{eqn:unin_pow_opt_3}\\
\text{s.t.}
&\qquad 0 \le \eta \le a_{i,j}, \,\forall j \in \mathcal{Q}_{s,d}, \IEEEyessubnumber \label{eqn:unin_pow_opt_31} \\
&\qquad \eqref{eqn:joint_opt_4}, \eqref{eqn:joint_opt_5}, \nonumber 
\end{IEEEeqnarray}
where $\eta$ is an auxiliary variable employed to facilitate the optimization. In this case, the objective is an affine function, and $a_{i,j}$ is a concave function for $\forall \, i,j \in \mathcal{Q}_{s,d}$.
It can easily be verified that the optimization problem in \eqref{eqn:unin_pow_opt_3} is convex, and therefore can be solved efficiently via interior point method using available standard optimization toolboxes (e.g., CVX \cite{grant2014cvx}). The overall alternating-optimization algorithm considering both the 3D trajectory and power allocation optimization is summarized in Algorithm~\ref{alg:alg}.
\begin{algorithm}
\caption{Proposed Alternating-Optimization Algorithm}
\begin{algorithmic}[t]
\State \textbf{Initialization:} Locations $\textbf{r}_i$ and power allocation $P_i$ for $i$th UAV for $\forall i \,{\in}\,\mathcal{N}\,{\cup}\,\mathcal{M}$, error tolerance $\varepsilon$
\State $t \gets 1$, $\textrm{R}_{\textrm{s} \rightarrow \textrm{d}}({-}1) \gets {-}\infty$, $\textrm{R}_{\textrm{s} \rightarrow \textrm{d}}(0) \gets 0$
\While{ $|\textrm{R}_{\textrm{s} \rightarrow \textrm{d}}(t{-}1) \,{-}\, \textrm{R}_{\textrm{s} \rightarrow \textrm{d}}(t{-}2)| \,{>}\, \varepsilon$ }
\State Compute $\frac{\partial \lambda_2({\mathcal{L}_\mathbf{W}})}{\partial x_i(t)}$, $\frac{\partial \lambda_2({\mathcal{L}_\mathbf{W}})}{\partial y_i(t)}$, $\frac{\partial \lambda_2({\mathcal{L}_\mathbf{W}})}{\partial z_i(t)}$ by \eqref{eq:spatial_gradient_unintended_1}-\eqref{eq:partial_derivative}
\State Update $\textbf{r}_i(t)$ in $\mathbb{R}^3$ as per \eqref{eq:loc_update_x} 
\State Compute optimal $P_i(t)$ by \eqref{eqn:unin_pow_opt_2} via CVX \cite{grant2014cvx} 
\State Compute $\textrm{R}_{\textrm{s} \rightarrow \textrm{d}}(t)$ by  Ford-Fulkerson algorithm~\cite{ford2015flows}
\State $t \gets t + 1$
\EndWhile
\end{algorithmic}\label{alg:alg}
\end{algorithm}

\section{Simulation Results} \label{sec:results}

\noindent In this section, we present numerical results based on extensive simulations, to evaluate the performance of the proposed joint 3D/2D trajectory and power allocation optimization. In our simulation environment, the BS and the UE are assumed to be located at $(0,0,h^\textrm{BS})$ and $(200\,\text{m},0,h^\textrm{UE})$, respectively, in $\mathbb{R}^3$ with $h^\textrm{BS} \,{=}\,15\,\text{m}$. Moreover, the SIs are located randomly in $xy-$plane with fixed altitude of $h^\textrm{SI} \,{=}\,20\,\text{m}$. The list of simulation parameters are given in Table~\ref{tab:simulation}. 
\begin{table}[h]
\caption {Simulation Parameters} \label{tab:simulation} 
\renewcommand{\arraystretch}{1.2}
\centering
\begin{tabular}{ lc }
\hline
Parameter & Value \\
\hline
\hline
Number of UAVs & 8\\
Number of SI Nodes & 7\\
Path-loss exponents $(\alpha_1, \alpha_2)$ & $2.05, 2.32$  \\
Maximum transmit power of the UAVs $(\textrm{P}_{\textrm{max}})$ & $20\,\text{dBm}$\\
Transmit power of the SI nodes $(\textrm{P}_j^\textrm{J}$, $\forall j\,{\in}\,\mathcal{M})$ & $30\,\text{dBm}$ \\
Bandwidth $({B})$ & $10\,\text{KHz}$ \\ 
Interference threshold $(\textrm{I}^{\textrm{th}}_j$, $\forall j\,{\in}\,\mathcal{M})$ & $[{-}50,{-}10]\,\text{dBm}$ \\ 
Interference radius $(r_{\textrm{int}})$ & $5\,\text{m}$ \\ 
Carrier frequency $(f_{\rm c})$ & $2~ \text{GHz}$ \\ 
Smoothed step-function parameters $(\zeta,\kappa,{\color{black}y_0})$ & $1,10, 10^{-3}$ \\
Safety precaution priority $(\chi)$ & $1$ \\
\hline
\end{tabular}
\end {table}

\begin{figure}[!t]
	\centering
	\includegraphics[width=0.47\textwidth]{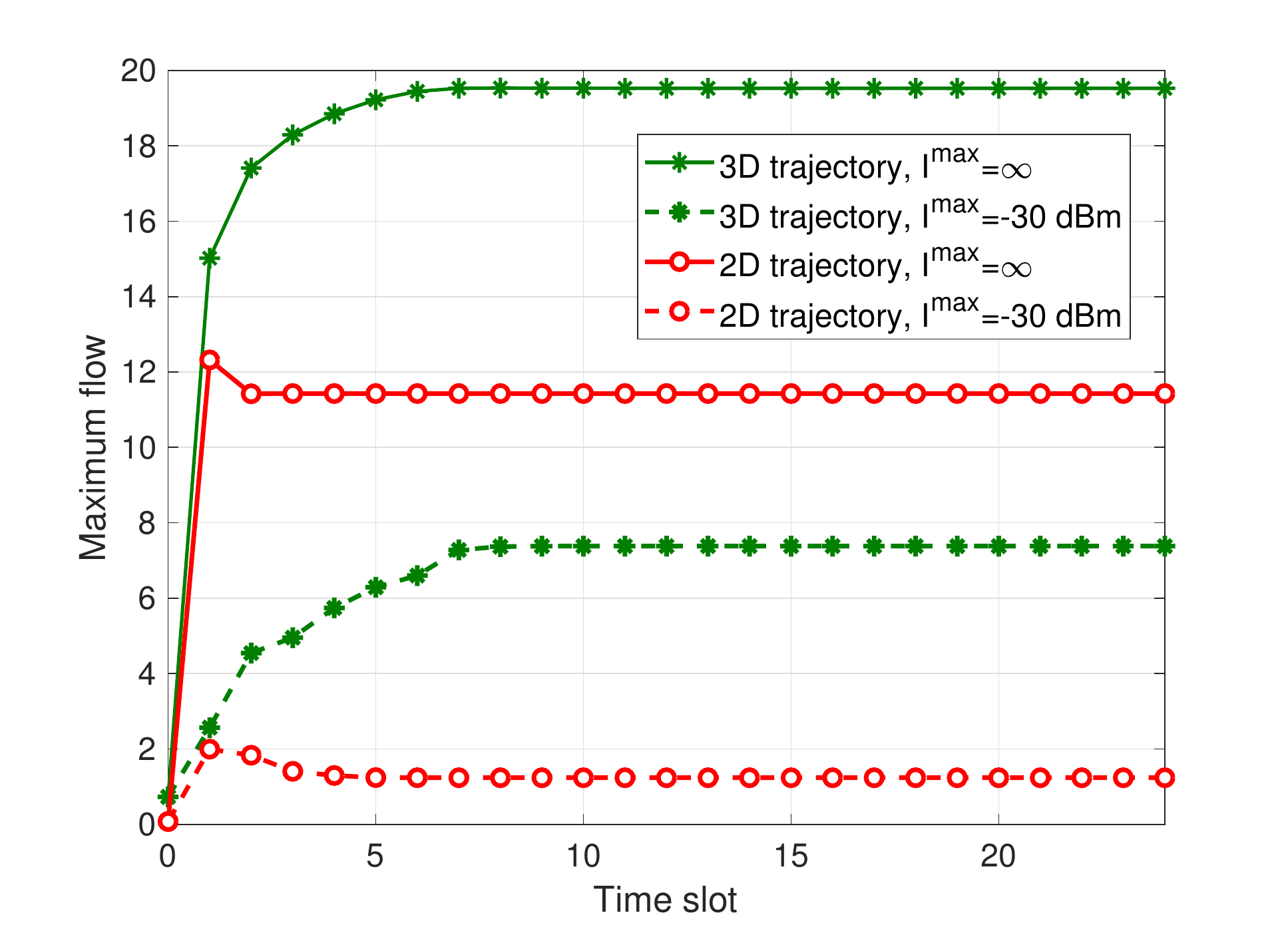}
		\vspace{-3mm}

	\caption{Convergence of the proposed joint power allocation and 3D/2D design.}
	\label{fig:flow_vs_iteration}
\end{figure}

In Fig.~\ref{fig:flow_vs_iteration}, for $h^\textrm{UE} \,{=}\,25\,\text{m}$ the convergence of the algorithm for different interference threshold is presented while considering 2D and 3D trajectory design algorithms. It can be seen that the 3D trajectory design needs more time for convergence while 2D trajectory deign converges faster. However, 3D trajectory design can perform  better than 2D and can double the transmission flow of the network. This can be a strong benefit of 3D trajectory design.
\begin{figure}[!t]
	\centering
	\includegraphics[width=0.47\textwidth]{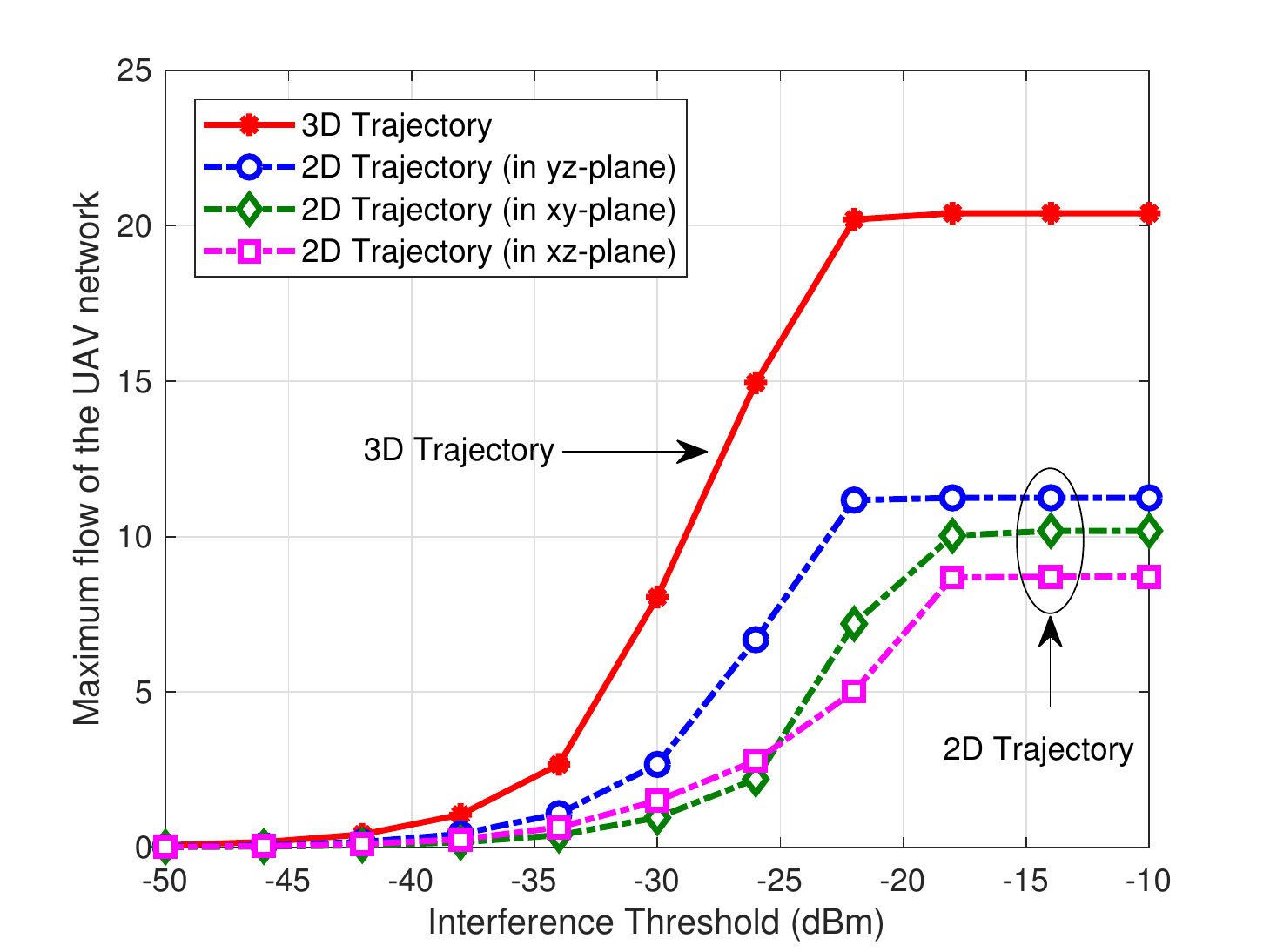}
		\vspace{-3mm}

	\caption{Maximum flow versus interference threshold for 2D and 3D trajectory optimization strategies along with optimal power allocation.}
	\label{fig:flow_vs_threshold}
\end{figure}

In Fig.~\ref{fig:flow_vs_threshold}, we depict the maximum flow against the interference threshold $\textrm{I}^\textrm{max}$ for the UE altitude of $h^\textrm{UE} \,{=}\,25\,\text{m}$, which may well represent a low-flying UAV as the desired UE. We assume that $\textrm{I}^{\textrm{max}}_{j} $ is the same for each SI $j \in \mathcal{M}$. We observe that when the relaying UAVs are allowed to optimize their trajectories using the proposed 2D trajectory design (i.e., in $xy-$, $xz-$, or $yz-$ planes), the performances are always inferior to that of the proposed 3D trajectory optimization.  Note that the decreasing interference threshold corresponds to the situation that the SI receivers are more susceptible to the interference. Defining $\textrm{I}^\textrm{max}$ as the maximum interference, below which the maximum flow starts decreasing, the trajectory optimization in $yz-$plane (2D) and in 3D ends up with being more robust against increasing the interference threshold to $\textrm{I}^\textrm{max}\,{=}\,{-}22\,\text{dBm}$, while the other 2D schemes have $\textrm{I}^\textrm{max}\,{=}\,{-}18\,\text{dBm}$. It should be noted that this observation depends on the initial locations of the UAVs as well.
\begin{figure}[!t]
	\centering
	\includegraphics[width=0.47\textwidth]{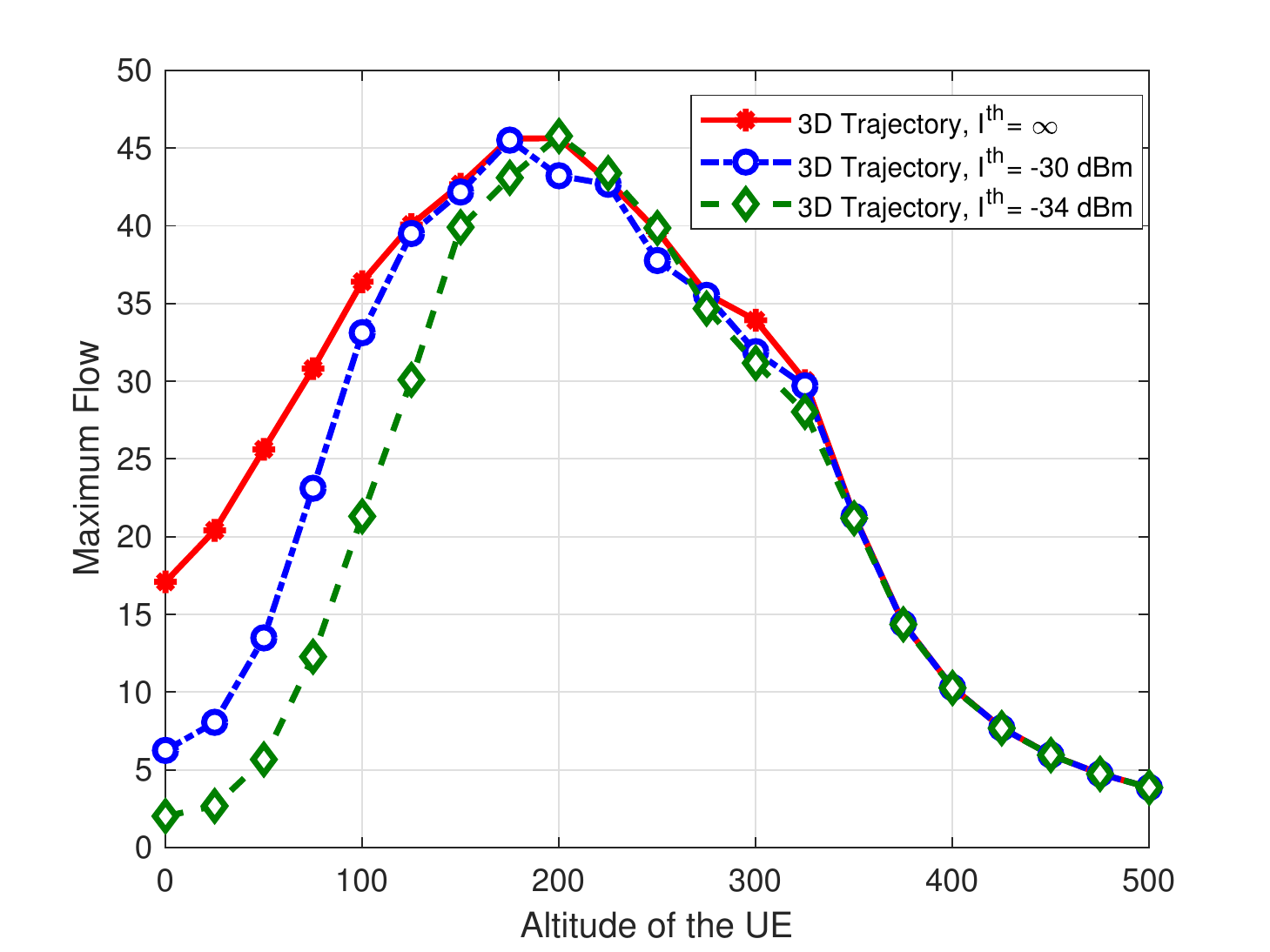}
	\vspace{-2mm}
	\caption{Maximum flow versus UE altitude for interference threshold of $\{{-}34,{-}30\}\,\text{dBm}$ as well as no interference threshold (i.e., $\infty$) with 3D trajectory design. 
	\label{fig:flow_vs_altitude}}

\end{figure}
In Fig.~\ref{fig:flow_vs_altitude}, we depict the maximum flow along with varying UE altitude of $h^\textrm{UE} \,{\in}\, [0,500]\,\text{m}$, which covers both on-ground and flying UEs. Interestingly, the maximum flow improves as the UE altitude increases till $h^\textrm{UE} \,{=}\, 200\,\text{m}$.  To illustrate this situation, we depict the respective 3D trajectories of all the UAVs in Fig.~\ref{fig:trajectory} for the UE altitudes of $h^\textrm{UE} \,{=}\, \{25,200\}\,\text{m}$. Moreover, the decrease in the maximum flow after $h^\textrm{UE} \,{=}\, 200\,\text{m}$ is mainly due to the increased path loss, which now becomes more dominant compared to the interference (even though the interference is also decreasing due to the increasing distance).  

\begin{figure}[!t]
	\centering
	\includegraphics[width=0.47\textwidth]{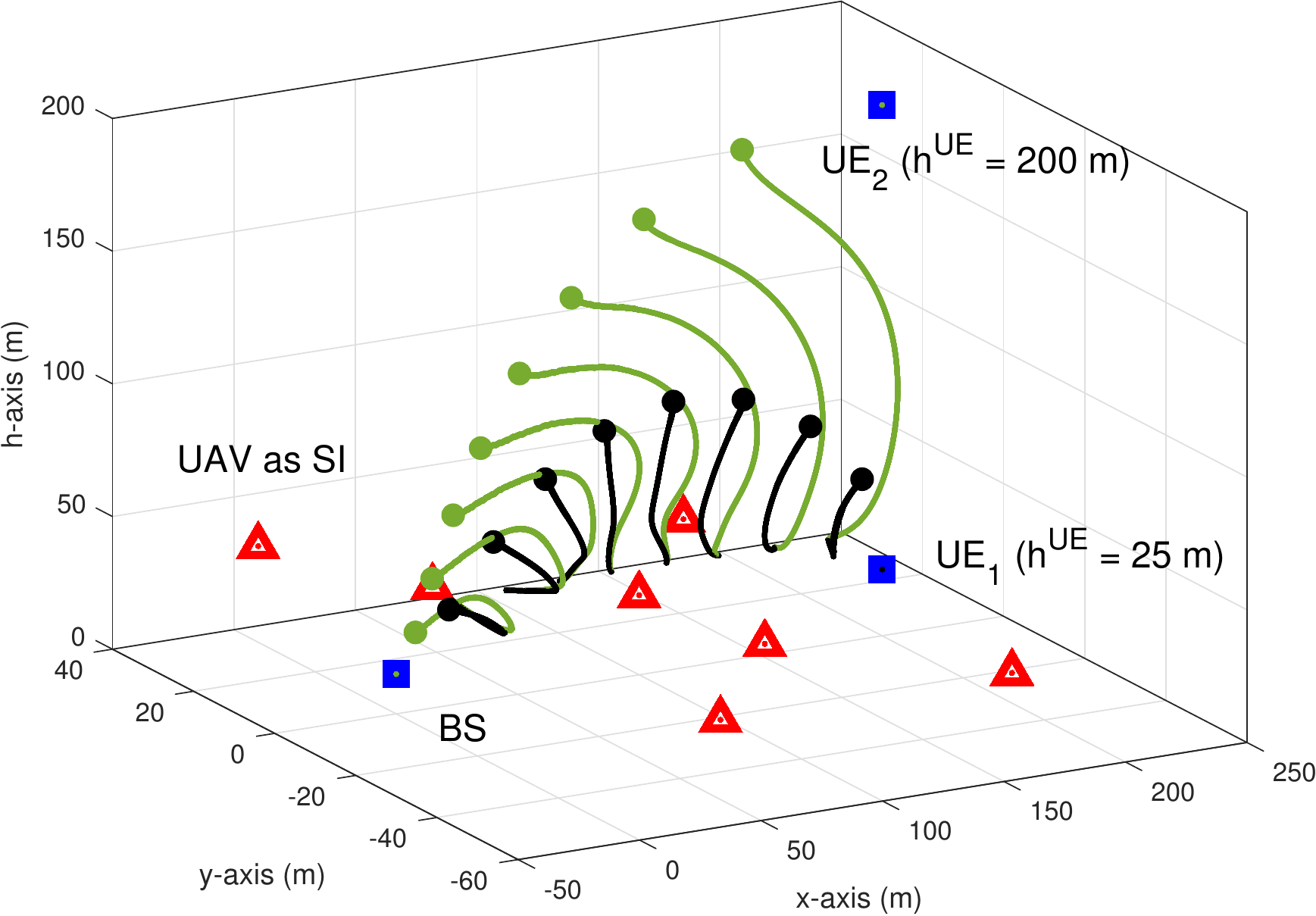}
		\vspace{-3mm}
	\caption{3D trajectories for the results of Fig.~\ref{fig:flow_vs_altitude} at UE altitudes of $\{25,200\}\,\text{m}$ (black trajectory is for 25 m and green trajectory is for 200 m.)}
	\label{fig:trajectory}
	\vspace{-0.2in}
\end{figure}

\section{Conclusion}\label{sec:conclusion}
\noindent In this paper, we considered the joint power and 3D trajectory design for a relay assisted UAV network coexisting with the already existing network. A joint optimization solution for 3D trajectory design and power allocation is proposed based on spectral graph theory and convex optimization. Simulation results demonstrate the effectiveness of the proposed algorithm in improving the maximum flow and mitigation of the unwanted interference to the co-existed network.

\bibliographystyle{IEEEtran}
\vspace{-3mm}
\bibliography{IEEEabrv,references}

\begin{thebibliography}{10}
\providecommand{\url}[1]{#1}
\csname url@samestyle\endcsname
\providecommand{\newblock}{\relax}
\providecommand{\bibinfo}[2]{#2}
\providecommand{\BIBentrySTDinterwordspacing}{\spaceskip=0pt\relax}
\providecommand{\BIBentryALTinterwordstretchfactor}{4}
\providecommand{\BIBentryALTinterwordspacing}{\spaceskip=\fontdimen2\font plus
\BIBentryALTinterwordstretchfactor\fontdimen3\font minus
  \fontdimen4\font\relax}
\providecommand{\BIBforeignlanguage}[2]{{%
\expandafter\ifx\csname l@#1\endcsname\relax
\typeout{** WARNING: IEEEtran.bst: No hyphenation pattern has been}%
\typeout{** loaded for the language `#1'. Using the pattern for}%
\typeout{** the default language instead.}%
\else
\language=\csname l@#1\endcsname
\fi
#2}}
\providecommand{\BIBdecl}{\relax}
\BIBdecl

\bibitem{hayat2016survey}
S.~Hayat, E.~Yanmaz, and R.~Muzaffar, ``Survey on unmanned aerial vehicle
  networks for civil applications: A communications viewpoint,'' \emph{Commun.
  Surveys Tuts.}, vol.~18, no.~4, pp. 2624--2661, 2016.

\bibitem{rahmati2019energy}
A.~Rahmati, Y.~Yap{\i}c{\i}, N.~Rupasinghe, I.~G\"{u}ven\c{c}, H.~Dai, and
  A.~Bhuyan, ``{Energy Efficiency of RSMA and NOMA in Cellular-Connected mmWave
  UAV Networks},'' in \emph{Proc. {IEEE} Int. Conf. Commun. (ICC) Workshops},
  Shanghai, China, May 2019.

\bibitem{mag}
Y.~Zeng, R.~Zhang, and T.~J. Lim, ``Wireless communications with unmanned
  aerial vehicles: Opportunities and challenges,'' \emph{IEEE Commun. Mag.},
  vol.~54, no.~5, pp. 36--42, 2016.

\bibitem{7577063}
A.~{Jaziri}, R.~{Nasri}, and T.~{Chahed}, ``Congestion mitigation in 5{G}
  networks using drone relays,'' in \emph{Proc. Int. Wireless Commun. Mobile
  Comput. Conf. (IWCMC)}, Paphos, Cyprus, Sep. 2016.

\bibitem{mag2}
H.~{Baek} and J.~{Lim}, ``Design of future {UAV}-relay tactical data link for
  reliable {UAV} control and situational awareness,'' \emph{IEEE Commun. Mag.},
  vol.~56, no.~10, pp. 144--150, 2018.

\bibitem{YOUNIS2008621}
M.~Younis and K.~Akkaya, ``Strategies and techniques for node placement in
  wireless sensor networks: A survey,'' \emph{Ad Hoc Networks}, vol.~6, no.~4,
  pp. 621 -- 655, 2008.

\bibitem{zhan2006wireless}
P.~{Zhan}, K.~{Yu}, and A.~{Lee Swindlehurst}, ``Wireless relay communications
  using an unmanned aerial vehicle,'' in \emph{Proc. IEEE 7th Workshop on
  Signal Processing Advances in Wireless Commun.}, Cannes, France, July 2006.

\bibitem{8424236}
Y.~{Chen}, N.~{Zhao}, Z.~{Ding}, and M.~{Alouini}, ``Multiple {UAVs} as relays:
  {Multi-Hop} single link versus multiple dual-hop links,'' \emph{IEEE Trans.
  Wireless Commun.}, vol.~17, no.~9, pp. 6348--6359, Sep. 2018.

\bibitem{faqir2018energy}
O.~J. Faqir, Y.~Nie, E.~C. Kerrigan, and D.~G{\"u}nd{\"u}z, ``Energy-efficient
  communication in mobile aerial relay-assisted networks using predictive
  control,'' \emph{IFAC-PapersOnLine}, vol.~51, no.~20, pp. 197--202, 2018.

\bibitem{8116613}
Y.~{Chen}, W.~{Feng}, and G.~{Zheng}, ``Optimum placement of {UAV} as relays,''
  \emph{IEEE Commun. Lett.}, vol.~22, no.~2, pp. 248--251, Feb 2018.

\bibitem{zhang2018joint}
S.~Zhang, H.~Zhang, Q.~He, K.~Bian, and L.~Song, ``Joint trajectory and power
  optimization for {UAV} relay networks,'' \emph{IEEE Commun. Lett.}, vol.~22,
  no.~1, pp. 161--164, 2018.

\bibitem{zeng2016throughput}
Y.~Zeng, R.~Zhang, and T.~J. Lim, ``Throughput maximization for {UAV}-enabled
  mobile relaying systems,'' \emph{IEEE Trans. on Commun.}, vol.~64, no.~12,
  pp. 4983--4996, 2016.

\bibitem{hosseinalipour2019interference}
S.~Hosseinalipour, A.~Rahmati, and H.~Dai, ``{Interference Avoidance Position
  Planning in {UAV}-assisted Wireless Communication},'' in \emph{Proc. {IEEE}
  Int. Conf. Commun. (ICC)}, Shanghai, China, May 2019.

\bibitem{bezdek2002some}
J.~C. Bezdek and R.~J. Hathaway, ``Some notes on alternating optimization,'' in
  \emph{AFSS Int. Conf. on Fuzzy Sys.}\hskip 1em plus 0.5em minus 0.4em\relax
  Springer, 2002, pp. 288--300.

\bibitem{tavana2017cooperative}
M.~Tavana, A.~Rahmati, V.~Shah-Mansouri, and B.~Maham, ``Cooperative sensing
  with joint energy and correlation detection in cognitive radio networks,''
  \emph{IEEE Commun. Lett.}, vol.~21, no.~1, pp. 132--135, 2017.

\bibitem{ahmed2016importance}
N.~Ahmed, S.~S. Kanhere, and S.~Jha, ``On the importance of link
  characterization for aerial wireless sensor networks,'' \emph{IEEE Commun.
  Mag.}, vol.~54, no.~5, pp. 52--57, 2016.

\bibitem{weibel2004safety}
R.~Weibel and R.~J. Hansman, ``Safety considerations for operation of different
  classes of {UAV}s in the {NAS},'' in \emph{AIAA 4th Aviation Technology,
  Integration and Operations (ATIO) Forum}, 2004, p. 6244.

\bibitem{6807812}
X.~He, H.~Dai, and P.~Ning, ``Dynamic adaptive anti-jamming via controlled
  mobility,'' \emph{IEEE Trans. Wireless Commun.}, vol.~13, no.~8, pp.
  4374--4388, Aug 2014.

\bibitem{chung1997spectral}
F.~R. Chung and F.~C. Graham, \emph{Spectral graph theory}.\hskip 1em plus
  0.5em minus 0.4em\relax American Mathematical Soc., 1997, no.~92.

\bibitem{rahmati2019dynamic}
A.~Rahmati, X.~He, I.~Guvenc, and H.~Dai, ``{Dynamic Mobility-Aware
  Interference Avoidance for Aerial Base Stations in Cognitive Radio
  Networks},'' in \emph{in Proc. IEEE INFOCOM}, Paris, France, 2019.

\bibitem{ford2015flows}
L.~R. Ford~Jr and D.~R. Fulkerson, \emph{Flows in networks}.\hskip 1em plus
  0.5em minus 0.4em\relax Princeton University Press, 2015.

\bibitem{grant2014cvx}
M.~Grant and S.~Boyd, ``{CVX}: {Matlab} software for disciplined convex
  programming, version 2.1,'' 2014.

\end{thebibliography}

\end{document}